\documentclass[epj,twocolumn]{webofc}
\usepackage[varg]{txfonts}   
\woctitle{Hadron Collider Physics symposium 2012}
\begin{document}
\title{Search For New Physics at BABAR}
%
%

\author{Romulus Godang
\inst{University~of~South~Alabama}
\fnsep
\thanks{
This work was supported by the U.S. Department of Energy under 
grant No. DE-FG02-96ER-40970
\email{godang@usouthal.edu}} 
}

\institute{Department of Physics \\
University of South Alabama      \\
ILB 115, 307 University Blvd., N.\\
\vspace{0.1cm}                   \\ 
SLAC-PUB-15354                   \\
USA-HEP-2013-02                  \\
UMS-HEP-2013-02                  
} 

\abstract
{%
Using a full BABAR data sample of 426 $fb^{-1}$, we present improved measurements of the ratio 
${\cal{R}}(D^{(*)}) = {\cal{B}} (\bar{B} \to D^{(*)}\tau^{-}\bar{\nu}_{\tau})/$
${\cal{B}} (\bar{B} \to D^{(*)}\ell_{\ell}^{-}\bar{\nu}_{\ell})$, where $\ell$ is 
either electron or muon. We measure ${\cal{R}}(D) = 0.440 \pm 0.058 \pm 0.042$ and
${\cal{R}}(D^*) = 0.332 \pm 0.024 \pm 0.018$. These ratios exceed the Standard Model
predictions by $2.0\sigma$ and $2.7\sigma$, respectively. The results disagree with 
the Standard Model predictions at the level of $3.4\sigma$. The ratios are sensitive to new
physics contributions in the form of a charged Higgs boson. However, the access cannot be
explained by a charged Higgs boson in the type II two-Higgs-doublet model.
}
\maketitle
\section{INTRODUCTION}
\label{intro}
The semileptonic physics in $B$ meson sector played a prominent role in investigating
of new physics effect at low-energy region. Semileptonic transitions are the simplest process 
in $B$ mesons decay. In the Standard Model (SM), the heavy $b$ quark decays to either a $c$ or an $u$ 
quark and the virtual $W$ boson~\cite{semi1, semi2, semi3}.  Experimentally, semileptonic decays have 
the advantage of large branching fraction and are used to determine the weak couplings, 
the Cabibbo-Kobayashi-Maskawa (CKM) matrix elements, $|V_{cb}|$ and $|V_{ub}|$~\cite{PDG12}. 

\begin{figure}[!htb]
\centering
\hspace*{-0.2cm}
\includegraphics[width=8cm,clip]{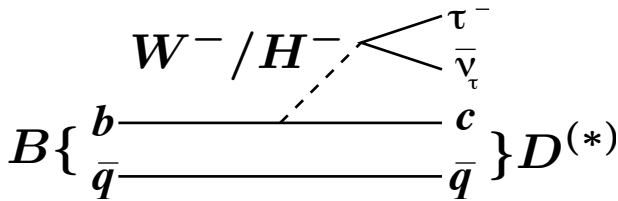}
\caption{The parton level diagram for $\bar{B} \to D^{(*)}\tau^{-}\bar{\nu}_{\tau}$ decays}
\label{Intro_Feynman}
\end{figure}

The parton level diagram of $\bar{B} \to D^{(*)}\tau^{-}\bar{\nu}_{\tau}$ decays 
where $D^{(*)}$ refers to either a $D$  or a $D^*$ meson is shown in 
Fig.~\ref{Intro_Feynman}. The decay $\bar{B} \to D^{(*)}\tau^{-}\bar{\nu}_{\tau}$ 
is sensitive to charged Higgs contribution at the tree level.
The three-body decay $\bar{B} \to D^{(*)}\tau^{-}\bar{\nu}_{\tau}$ permits the study of decay
distribution which discriminate between $W^-$ and $H^-$ exchange~\cite{higgs1, higgs2}. 

The decay of $\bar{B} \to D^{(*)}\tau^{-}\bar{\nu}_{\tau}$ with $\tau$ lepton in the final state
offer possibilities of significant new physics contributions that is not present in the process
where light lepton such as electron and muon in the final state. The study of 
$\bar{B} \to D^{(*)}\tau^{-}\bar{\nu}_{\tau}$ have already shown the new physics
contributions can be over-constrained~\cite{higgs3, higgs4, higgs5, higgs6}.  
The existing studies of the $\bar{B} \to D^{(*)}\tau^{-}\bar{\nu}_{\tau}$ based on two-Higgs-doublet
model (2HDM) predict a substantial impact on the ratio ${\cal{R}}(D^{(*)})$ and ${\cal{R}}(D)$
~\cite{higgs3, higgs4, higgs6, higgs7}.

\section{The BABAR DETECTOR AND DATA SET}
\label{sec-1}
The BABAR detector was operated at the PEP-II asymmetric-energy storage rings at
the SLAC National Accelerator Laboratory. The data used in this analysis were 
collected with the BABAR detector. We analyze  data recorded with the BABAR
detector at a center of mass energy of 10.58 GeV. The data sample consist of an 
integrated luminosity of 426 $fb^{-1}$, corresponding to 471 $\times 10^6 B\bar{B}$ pairs.
An additional sample of 40 $fb^{-1}$, taken at energy 40 MeV below the $\Upsilon(4S)$ 
resonance. This additional sample of data is used to study the continuum background from 
the decays of $e^+ e^-  \to q \bar{q} (\gamma)$ pairs where $q$ can be $u, d, s, c, \tau$. 

A detail description of the BABAR detector is presented elsewhere~\cite{Babar_nim}.
The momenta of the charged particles are measured in a tracking system consisting 
of a 5-layer double sided silicon vertex tracker (SVT) and a 40-layer drift
chamber (DCH). The SVT and DCH operate within a 1.5 T solenoid field and have a combined 
solid angle coverage in the center of mass frame of 90.5\%. A detector of internally reflected
Cerenkov radiation (DIRC) is used for charged particle identifications of pions, kaons, 
and protons with likelihood ratios calculated from $dE/dx$ measurements in the SVT and DCH.  
Photons and long-lived neutral hadrons are detected and their energies are measured in 
a CsI(Tl) electromagnetic calorimeter (EMC). For electrons, energy lost due to 
bremsstrahlung is recovered from deposits in the EMC.

\section{ANALYSIS}
In this analysis, instead of measuring the absolute branching fraction of 
$\bar{B} \to D^{(*)}\tau^{-}\bar{\nu}_{\tau}$, we measure the ratios
\begin{equation}
{\cal{R}}(D^{(*)}) = \frac
{{\cal{B}} (\bar{B} \to D^{(*)}\tau^{-}\bar{\nu}_{\tau})}
{{\cal{B}} (\bar{B} \to D^{(*)}\ell^{-}\bar{\nu}_{\ell})}
\end{equation}
where $\ell$ is either electron or muon. In the standard model (SM), the relative rate
${\cal{R}}(D^{(*)})$ have less than 6\% uncertainty~\cite{higgs7}.
In the decay of $\bar{B} \to D^{(*)}\tau^{-}\bar{\nu}_{\tau}$, we construct the $\tau$ 
lepton only from the purely lepton decays: $\tau^- \to e^- \bar{\mu}_e \nu_{\tau}$ 
and $\tau^- \to e^-\bar{\nu}_{\mu} \nu_{\tau}$ so that the signal events 
$(\bar{B} \to D^{(*)}\tau^{-}\bar{\nu}_{\tau})$ and 
the normalization events $(\bar{B} \to D^{(*)}\ell^{-}\bar{\nu}_{\ell})$ are 
identified by the same particles in the final state. When taking the ratio of 
${\cal{R}}(D^{(*)})$, the various sources of uncertainties will be canceled and
reduced.

We reconstruct candidate events produced in $\Upsilon(4S) \to B\bar{B}$ decays by selecting
the hadron decay of one of the $B$ meson ($B_{tag}$). The other candidate events are 
reconstructed semileptonically, specially a charm meson (either charged or neutral $D$ or
$D^*$) and a charged lepton (either $e$ or $\mu$). The signal events 
$(\bar{B} \to D^{(*)}\tau^{-}\bar{\nu}_{\tau})$ and the normalization events
$(\bar{B} \to D^{(*)}\ell^{-}\bar{\nu}_{\ell})$ are extracted using unbinned 
maximum-likelihood fit to the two-dimensional distributions of the invariant mass
of the undetected particles. Basically it is the invariant mass of the neutrinos.
\begin{equation}
m^2_{miss} = p^2_{miss} = (p_{e^+e^-} -p_{B_{tag}} - p_{D^{(*)}} -p_{\ell})^2
\end{equation}  
where $p$ is the four-momenta of the colliding beams, $B_{tag}$, $D^{(*)}$ and charged
lepton, respectively. The lepton three-momentum in the $B$ rest frame is denoted 
by $\boldmath{p}^*_{\ell}$.  The distribution of the lepton three-momentum of the 
signal events is softer than the distribution of the lepton three-momentum of the normalization
events because the observed lepton in the signal events is a secondary particle originated from
the $\tau$ decay, $\tau^- \to \ell^- \bar{\nu}_\ell \nu_{\tau}$.

If all particles are properly reconstructed the invariant mass
of the undetected particles ($m^2_{miss}$) with a single missing neutrino peaks at zero, whereas
the signal events which have three missing neutrinos have a wide $m^2_{miss}$ distribution
that extends from -1 GeV$^2$ to 10 GeV$^2$. The two observable kinematic variables are used to select 
the $B_{tag}$ candidates:
\begin{equation}
m_{ES} = \sqrt{E^2_{beam} - \boldmath{p}^2_{tag}}
\end{equation} 
and
\begin{equation}
\Delta E = E_{tag} - E_{beam}
\end{equation} 
where the $\boldmath{p}_{tag}$ and $E_{tag}$ refer to the center-of-mass momentum and 
energy of the $B_{tag}$. $E_{beam}$ is the center-of-mass of a single beam particle.
If the $B$ decays are correctly reconstructed, the distribution of the $m_{ES}$ is 
centered at the $B$ meson mass with a resolution of 2.5 MeV. The distribution 
of $\Delta E$ is centered at zero with a resolution of 18 MeV.  
In this analysis we required $m_{ES}$ > 5.27 GeV and $|\Delta E|$ < 0.072 GeV.

\begin{figure}[!htb]
\centering
\hspace*{-0.2cm}
\includegraphics[width=8.4cm,clip]{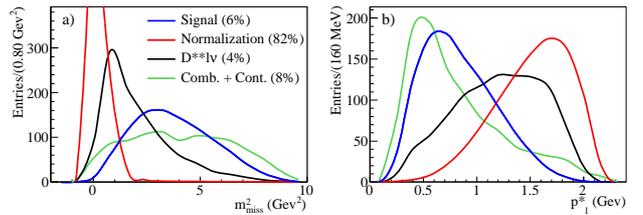}
\caption{The comparison of the data and the fit projections for 
the four $D^{(*)}\ell$ samples: the signal events (6\%), normalization events (82\%), 
$B \to D^{**} \ell \nu$ (4\%), and combinatoric and continuum background (8\%). The 
$m_{ES}$ is on the left and the $p^*_{\ell}$ is on the right. All distributions are 
normalized to 1000 entries after all selections are applied.} 
\label{Intro_M2Pl}
\end{figure}

The main background contributions to the signal events are the following:
\begin{itemize}
\item The decay of $\bar{B} \to D^{**}(\tau^-/\ell^-)\bar{\nu}_{\ell}$ where $D^{**}$ 
mesons refer to the charm resonances heavier than the $D^*$ meson such as
$D^*_0$, $D_1$, $D'_1$, and $D^*_2$ orbital excitations of the $c\bar{q}$ pairs. 
The decay of $\bar{B} \to D^{**}\ell^-\bar{\nu}_{\ell}$ where the $D^*$ meson
decays to $D^{(*)}\pi^0$ peaks in the $m^2_{miss}$ distribution. These events are
estimated using the Monte Carlo samples.
\item Charge cross-feed events: these events come from the decay of 
$\bar{B} \to D^{**}(\tau^-/\ell^-)\bar{\nu}_{\ell}$. These background events were
reconstructed with the incorrect charge where one of the charges particles in 
the final state has been assigned to the different $B$ meson.
\item Other $B\bar{B}$ background: These events come from the decay of 
$B \to D^{(*,**)}D_s^{(*,**)+}$ due to the large leptonic and semileptonic
branching fractions of $D^+_s$ mesons. We estimate these events using Monte
Carlo sample and its contribution is fixed in the fitting process.
\item Continuum background: these events come from the decay of $e^+ e^-  
\to q \bar{q} (\gamma)$ pairs where $q$ can be $u, d, s, c, \tau$. To estimate this
background we use the additional data sample of 40 $fb^{-1}$, taken at energy 40 MeV 
below the $\Upsilon(4S)$ resonance.
\end{itemize}
Figure~\ref{Intro_M2Pl} shows the distributions of $m^2_{miss}$ and $p^*_{\ell}$
variables for the signal events (6\%), normalization events (82\%), $B \to D^{**}
\ell \nu$ (4\%), and combinatoric and continuum background (8\%). All distributions are 
normalized to 1000 entries after all selections are applied. 

The results of the signal and normalization yields are extracted using unbinned 
maximum-likelihood fit to the two dimensional, $m^2_{miss}$-$\boldmath{p}^*_{\ell}$
contributions. The fit is performed simultaneously to the four 
$D^{(*)}\ell$ samples and four $D^{(*)}\pi^0\ell$ samples. The distributions of
each $D^{(*)}\ell$ and $D^{(*)}\pi^0\ell$ sample is fitted to the sum of either eight
or six contributions, respectively. The fit relies on $8 \times 4 + 6 \times 4 = 56$
probability density functions (PDFs). The two dimensional, 
$m^2_{miss}$-$\boldmath{p}^*_{\ell}$ contributions for each of the 56 PDFs are described
in detail using smooth non-parametric kernel estimator~\cite{estimator}. 

\begin{figure}[!htb]
\centering
\hspace*{-0.5cm}
\includegraphics[width=8.4cm,clip]{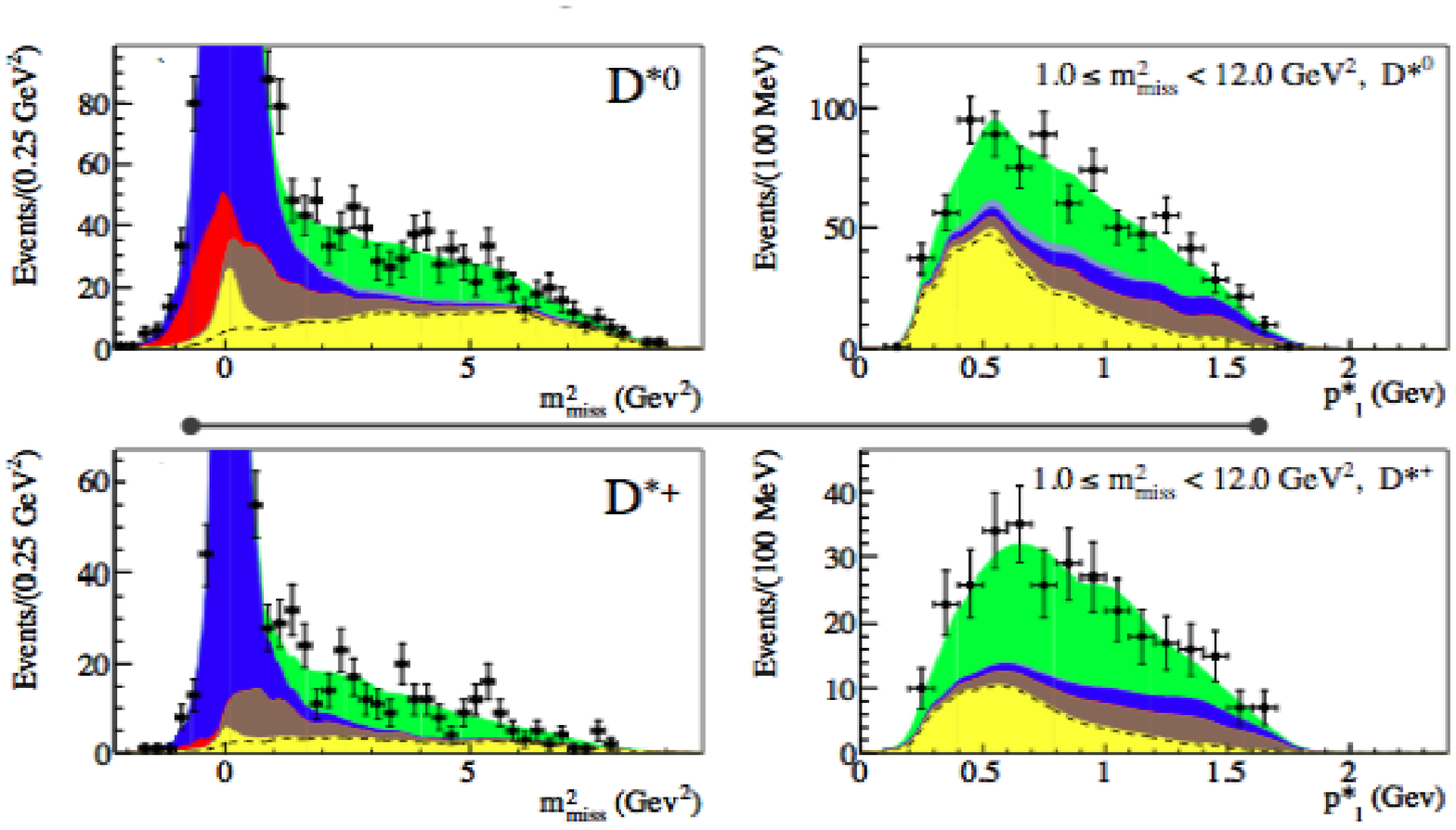}
\caption{The comparison of the $m^2_{miss}$ (left) and $\boldmath{p}^*_{\ell}$ (right) 
distributions of the $B \to D^* \tau \nu$ (data points) with the projections of the results of the 
isospin-unconstrained fit. The region above the dashed line is the background component 
corresponds to $B\bar{B}$ background. The region below the dashed line corresponds to 
the continuum background. In the $\boldmath{p}^*_{\ell}$ distributions we only include 
events with $m^2_{miss}> 1$ GeV$^2$.}
\label{Dstaunu_result}
\end{figure}

The $m^2_{miss}$ distributions of the signal events and the normalization events cane be
easily distinguished due to the different number of neutrino as the undetected particles
in its corresponding decays in the final state. However, the $m^2_{miss}$ distributions of the 
backgrounds resemble those of the signal events, and therefore in the fitting procedure
these contributions are either fixed fitted or constrained by the $D^{(*)}\pi^0\ell$ 
Monte Carlo sample.   
Figure~\ref{Dstaunu_result} shows the yield of $B \to D^* \tau \nu$ and the comparison of the 
$m^2_{miss}$ and $\boldmath{p}^*_{\ell}$ distributions of the $B \to D^* \tau \nu$ (data points) 
with the projections of the results of the isospin-unconstrained fit. The region above the dashed line
is the background component corresponds to $B\bar{B}$ background. 

The region below the dashed line
corresponds to the continuum background. Due to the charged cross-feed events the $m^2_{miss}$
distribution peak around $m^2_{miss}=0$ in the background component. However, in the 
$\boldmath{p}^*_{\ell}$ distributions, we only include events with $m^2_{miss}> 1$ GeV$^2$.

\begin{figure}[!htb]
\centering
\hspace*{-0.5cm}
\includegraphics[width=8.4cm,clip]{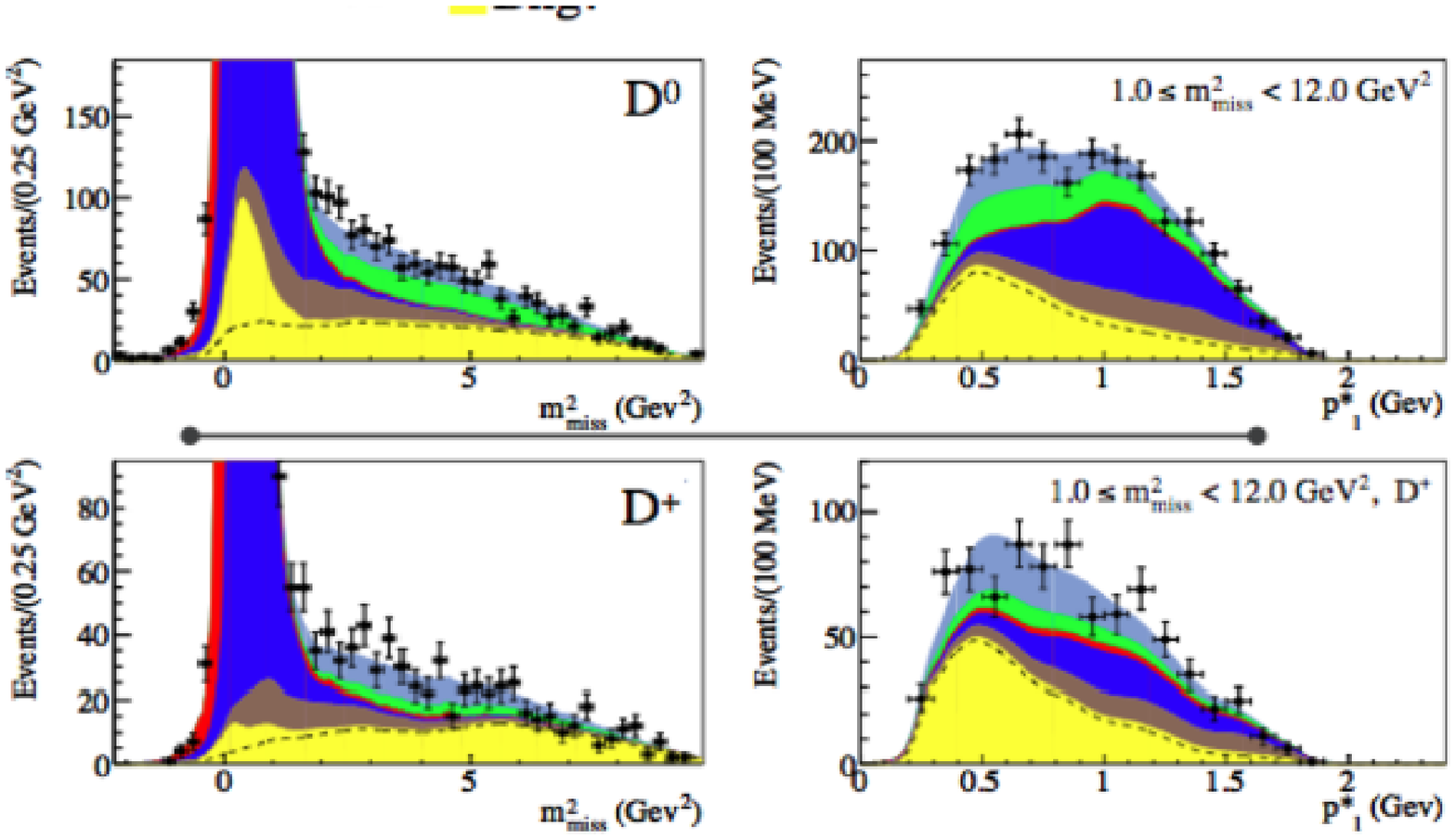}
\caption{The comparison of the $m^2_{miss}$ (left) and $\boldmath{p}^*_{\ell}$ (right) 
distributions of the $B \to D \tau \nu$ (data points). with the projections of the results of the 
isospin-unconstrained fit. The region above the dashed line is the background component 
corresponds to $B\bar{B}$ background. The region below the dashed line corresponds to 
the continuum background. In the $\boldmath{p}^*_{\ell}$ distributions we only include 
events with $m^2_{miss}> 1$ GeV$^2$.}
\label{Dtaunu_result}
\end{figure}

We extract the branching fraction ratios as define in the following 
\begin{equation}
{\cal{R}}(D^{(*)}) = \frac{N_{sig}}{N_{norm}} \times \frac{\epsilon_{sig}}{\epsilon_{norm}}
\end{equation} 
where $N_{sig}$ and $N_{norm}$ are the number of signal and normalization events extracting from 
the fitting process, respectively. The $\epsilon_{sig}/\epsilon_{norm}$ is the ratio of the 
efficiencies of the signal and the normalization events. We impose the isospin relations of
${\cal{R}}(D^*) \equiv {\cal{R}}(D^{*+}) = {\cal{R}}(D^{*0})$ and ${\cal{R}}(D) \equiv {\cal{R}}(D^+)
= {\cal{R}}(D^0)$.
Table~\ref{table1} shows the fit results of the yield of $B \to D^* \tau \nu$ with 
the statistical uncertainties only. 
\begin{table}
\centering
\caption{The yield results for the $B \to D^* \tau \nu$ channel where the uncertainties are the
statistical only.}
\label{table1}    
\begin{tabular}{llll}
\hline\hline
Mode               & $D^{*0} \tau \nu$ & $D^{*+} \tau \nu$  & $D^{*} \tau \nu$ \\\hline
$N_{sig}$          & $639 \pm 62$      & $245 \pm 27$       & $888 \pm 63$     \\
${\cal{R}}(D^*)$   & $0.32 \pm 0.03$   & $0.36 \pm 0.04$    & $0.33 \pm 0.02$  \\
${\cal{B}}(D^* \tau \nu)$ & $1.71 \pm 0.17$ & $1.74 \pm 0.19$ & $1.76 \pm 0.13$ \\\hline\hline 
\end{tabular}
\end{table}
Figure~\ref{Dtaunu_result} shows the yield of $B \to D \tau \nu$ comparison of the 
$m^2_{miss}$ (left) and $\boldmath{p}^*_{\ell}$ (right) distributions of the $B \to D \tau \nu$ 
(data points) with the projections of the results of the isospin-unconstrained fit. 

The region above the dashed line is the background component corresponds to $B\bar{B}$ background. 
The region below the dashed line corresponds to the continuum background. 
In the $\boldmath{p}^*_{\ell}$  distributions, we only include events with $m^2_{miss}> 1$ GeV$^2$.
Table~\ref{table2} shows the fit results of the yield of $B \to D \tau \nu$ with 
the statistical uncertainties only. 
\begin{table}
\centering
\caption{The yield results for the $B \to D \tau \nu$ channel where the uncertainties are the
statistical only.}
\label{table2}    
\begin{tabular}{llll}
\hline\hline
Mode               & $D^{0} \tau \nu$  & $D^{+} \tau \nu$   & $D \tau \nu$     \\\hline
$N_{sig}$          & $314 \pm 60$      & $177 \pm 31$       & $489 \pm 63$     \\
${\cal{R}}(D)$     & $0.43 \pm 0.08$   & $0.47 \pm 0.08$    & $0.44 \pm 0.06$   \\
${\cal{B}}(D \tau \nu)$ & $0.99 \pm 0.19$ & $1.01 \pm 0.18$ & $1.02 \pm 0.13$ \\\hline\hline 
\end{tabular}
\end{table}

\section{SYSTEMATICS UNCERTAINTIES}
The largest systematic uncertainties in this analysis is due to the poorly understood
of the decay $B \to D^{**}(\ell/\tau) \nu$ background. The systematic uncertainty due
the PDF that describe these contributions including the uncertainty on the branching 
fractions of the four $B \to D^{**}\ell\nu$ decays, the branching fraction ratio
of $B \to D^{**}\tau\nu$ to $B \to D^{**}\ell\nu$, and its relative efficiency. We assign
2.1\% on ${\cal{R}}(D)$ and 1.8\% on ${\cal{R}}(D^*)$, respectively. 

We also assign a systematic uncertainty due to the observed variation of the decay of 
$B \to D^{*}\eta\ell\nu$, non-resonance $B \to D^{*}\pi(\pi)\ell\nu$, $B \to D^{**}(\ell/\tau)\nu$, 
and $D^{**} \to D^{(*)} \pi\pi$. They are 2.1\% for ${\cal{R}}(D)$ and 2.6\% for ${\cal{R}}(D^*)$.

The other largest systematic uncertainties are due to the continuum and $B\bar{B}$ 
backgrounds. We assign 4.9\% for ${\cal{R}}(D)$ and 2.7\% for ${\cal{R}}(D^*)$.
The systematic uncertainties due to the PDFs for the signal and normalization decays 
are 4.3\% for ${\cal{R}}(D)$ and 2.1\% for ${\cal{R}}(D^*)$. The systematic uncertainties
due to the efficiency ratios $\epsilon_{sig}$ and $\epsilon_{norm}$ are 2.6\% on 
${\cal{R}}(D)$ and 1.6\% on ${\cal{R}}(D^*)$, respectively. 

By choosing the decay of $\tau$ lepton only from the purely lepton decays: 
$\tau^- \to e^- \bar{\nu}_e \nu_{\tau}$ and $\tau^- \to e^-\bar{\nu}_{\mu} \nu_{\tau}$, 
uncertainties due to the particle identification, final state radiation, soft-pion reconstruction, 
and other related detector performance are largely cancel in taking the ratios. They only 
contribute about 1\% in the systematic uncertainty. 

All systematic uncertainties are added 
in quadrature to assign the total systematic uncertainty. There is a positive correlation 
between some of the systematic uncertainties ${\cal{R}}(D)$ and ${\cal{R}}(D^*)$. 
As a result the correlation of the total uncertainties is reduced to -0.27 for 
${\cal{R}}(D)$ and ${\cal{R}}(D^*)$. 

\section{CONCLUSIONS}

We have measured the ${\cal{R}}(D) = 0.440 \pm 0.058 \pm 0.042$ and
${\cal{R}}(D^*) = 0.332 \pm 0.024 \pm 0.018$. These ratios exceed the Standard Model
predictions by $2.0\sigma$ and $2.7\sigma$, respectively. They are disagree with 
the SM prediction at the level of $3.4\sigma$. The results are compatible
with the results measured by the Belle Collaboration~\cite{belle1, belle2}. Together with 
the results measured by the Belle Collaboration, it could be an indication of new physics 
processes in $B$ mesons decay.

\begin{figure}[!htb]
\centering
\includegraphics[width=8.4cm,clip]{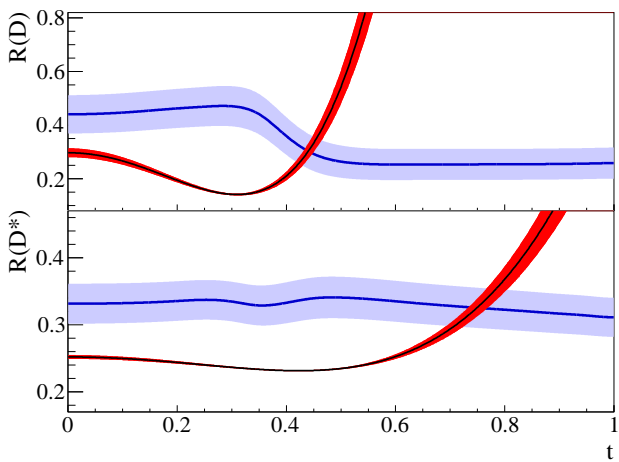}
\caption{The comparison results of this paper (blue band) with predictions that include
a charged Higgs boson of type II 2HDM (red band). The SM corresponds to 
$tan\beta/m_{H^+}= 0$.}
\label{PRL_Higgspsfrag}
\end{figure}

The measured values of ${\cal{R}}(D)$ and ${\cal{R}}(D^*)$ match the predictions of the 
particular Higgs model where the ratio of neutral Higgs field vacuum expectation values, 
$tan\beta$, and the mass of the physical charged Higgs boson, $m_{H^+}$ is $tan\beta/m_{H^+}= 
0.44 \pm 0.02$ GeV$^{-1}$ and $tan\beta/m_{H^+}= 0.75 \pm 0.04$ GeV$^{-1}$, respectively. 
Figure~\ref{PRL_Higgspsfrag} shows the comparison results of this paper 
with the charged Higgs boson of type II 2HDM predictions. The Standard Model 
predictions correspond to $tan\beta/m_{H^+} = 0$. 
However, these results are not compatible with a charged Higgs boson in the type II 
2HDM with a 99.8\% confidence level for any value of $tan\beta/m_{H^+}$.
More general charged-Higgs models or other New Physics contributions may explain
the access of ${\cal{R}}(D)$ and ${\cal{R}}(D^*)$.

\section{ACKNOWLEDGMENTS}

The author would like to thank the organizers of the Hadron Collider Physics 
Symposium 2012, Kyoto University, in Kyoto, Japan. The supports from the BABAR 
Collaboration, the University of South Alabama, and the University of Mississippi 
are gratefully acknowledged.

\end{document}